\documentclass[PRD,twocolumn,showpacs,superscriptaddress,preprintnumbers,nofootinbib,amsmath,amssymb]{revtex4}

\usepackage{graphicx}
\usepackage{dcolumn}
\usepackage{bm}

\usepackage{epstopdf}

\def\fnl{f_{\mathrm{nl}}}

\def\VEV#1{\left\langle #1 \right\rangle}
\def\Mpl{M_{\mathrm{ Pl}}}

\newcommand{\beq}{\begin{equation}}
\newcommand{\eeq}{\end{equation}}
\newcommand{\beqa}{\begin{eqnarray}}
\newcommand{\eeqa}{\end{eqnarray}}

\newcommand{\LL}{{\mathcal{L}}}

\newcommand{\bk}{{\mathbf{k}}}
\newcommand{\bx}{{\mathbf{x}}}
\newcommand{\bv}{{\mathbf{v}}}
\newcommand{\bw}{{\mathbf{w}}}
\newcommand{\bu}{{\mathbf{u}}}
\newcommand{\bq}{{\mathbf{q}}}
\newcommand{\bz}{{\mathbf{z}}}
\newcommand{\nn}{\nonumber}

\begin{document}

\preprint{IFT-UAM/CSIC-10-09, CERN-PH-TH/2010-039}
\title{Non-Gaussianity from Self-Ordering Scalar Fields}

\author{Daniel G.~Figueroa}
\affiliation{Department of Physics, CERN - Theory Division,
      CH-1211 Geneva 23, Switzerland}
\affiliation{Instituto de F\'isica Te\'orica UAM/CSIC,
    Universidad Aut\'onoma de Madrid, 28049 Madrid, Spain}
\author{Robert R.~Caldwell}
\affiliation{Department of Physics \& Astronomy, Dartmouth
    College, Hanover, NH 03755, USA}
\author{Marc Kamionkowski}
\affiliation{California Institute of Technology, Mail Code 350-17,
    Pasadena, California 91125, USA}

\date{\today}

\begin{abstract}

The Universe may harbor relics of the post-inflationary epoch in the form of a
network of self-ordered scalar fields. Such fossils, while consistent with
current cosmological data at trace levels, may leave too weak an imprint on the
cosmic microwave background and the large-scale distribution of matter to allow
for direct detection. The non-Gaussian statistics of the density perturbations
induced by these fields, however, permit a direct means to probe for
these relics. Here we calculate the bispectrum that arises in models of
self-ordered scalar fields. We find a compact analytic expression for the
bispectrum, evaluate it numerically, and provide a simple approximation that may
be useful for data analysis. The bispectrum is largest for triangles that are
aligned (have edges $k_1\simeq 2 k_2 \simeq 2 k_3$) as opposed to the
local-model bispectrum, which peaks for squeezed triangles ($k_1\simeq k_2 \gg
k_3$), and the equilateral bispectrum, which peaks at $k_1\simeq k_2 \simeq
k_3$.  We estimate that this non-Gaussianity should be detectable by the Planck
satellite if the contribution from self-ordering scalar fields to primordial
perturbations is near the current upper limit.

\end{abstract}


\maketitle

\section{Introduction}

A wealth of precise cosmological data are in good
agreement with the predictions of the simplest single-field
slow-roll (SFSR) inflationary models \cite{inflation}.  Still,
no theorist considers these as anything more than toy models.
Realistic models must surely be more complicated, and they
generically predict that there should arise, at some point,
observable phenomena that depart from the predictions of SFSR
inflation. Some possible directions for physics beyond
the SFSR approximation include multi-field models
\cite{larger,curvaton} and inflaton models with
non-standard kinetic terms \cite{Dvali:1998pa}.  There has also
been investigation of the consequences of topological defects \cite{topdefects} produced
toward the end of or after inflation \cite{endinflation}.

If inflation was followed by a transition associated
with the breaking of a global $O(N)$ symmetry, then
self-ordering scalar fields (SOSFs) are another possibly
observable early-Universe relic, even if there are no
topological defects (i.e., if $N > 4$).  Here, the alignment of
the scalar field as the Universe expands gives rise to a
scale-invariant spectrum of isocurvature perturbations,
without topological defects \cite{Turok:1991qq}.  Sample variance on the current
data limit these perturbations to
contribute no more than $\sim10\%$ of large-angle
cosmic-microwave-background (CMB) anisotropy power
\cite{Bevis:2004wk,Crotty:2003rz}. SOSF models are parametrized simply by the number $N$ of scalar
fields and the vacuum expectation value $v$.  The CMB constraint
implies $(v/N^{1/4}) \lesssim 5\times10^{15}$~GeV, as we
explain below. At this low amplitude, it is unlikely that any surviving relics
leave a distinct imprint on the CMB power spectrum \cite{CMBtexture}.

In recent years, non-Gaussianity has been developed as a novel
tool to investigate beyond-SFSR physics
\cite{NGReview,Bernardeau:2001qr}.  SFSR models do not predict
that primordial perturbations should be Gaussian, but
the departures from Gaussianity that they predict are
unobservably small \cite{localmodel,Wang:2000,Maldacena:2002vr}.
Multi-field models \cite{larger}, such as curvaton models
\cite{curvaton}, string-inspired DBI \cite{Dvali:1998pa,Equil}
models, and models with features in the inflaton potential
\cite{Wang:2000,Hannestad:2009yx} can
all produce larger, and possibly observable, deviations from
non-Gaussianity.  For example, the detailed shape (triangle dependence) of 
the bispectrum may also help distinguish these different
scenarios.  The ``local-model'' bispectrum, like
that which arises in curvaton and multi-field models, has a
very different shape dependence than ``equilateral-model''
bispectra, like those in DBI models.  Non-Gaussianity can be
sought in the CMB \cite{Luo:1993xx},
large-scale structure (LSS) \cite{LSS}, and the
abundances and properties of gravitationally-bound objects
\cite{abundances} or voids
\cite{Kamionkowski:2008sr}.  Biasing may significantly amplify
the effects of non-Gaussianity
\cite{Dalal:2007cu} in the galaxy
distribution.

The energy-density perturbations in self-ordering scalar
fields are quadratic in the scalar-field perturbation, which may
itself be approximated as a Gaussian field.  The density
perturbations induced by SOSFs are thus expected to be highly
non-Gaussian \cite{Turok:1991qq,Jaffe:1993tt,DefectBITRISPECTRA}, even in the
absence of topological defects.  It is thus
plausible that the non-Gaussianity induced by SOSFs might be
detectable, even if they provide only a secondary contribution
to primordial perturbations.

In this paper, we perform the first calculation of the full
shape (triangle) dependence of the bispectrum from SOSFs.  We
follow the formalism for non-Gaussianity developed in
Ref.~\cite{Jaffe:1993tt}.  We find considerably simplified
formulas for the bispectrum, evaluate them numerically, and find
a simple approximation to aid in data-analysis efforts.
We estimate the current non-Gaussianity constraint to the model
parameter space and find it to be comparable to that from the
upper limit to isocurvature perturbations from CMB fluctuations.

The plan of this paper is as follows:  In Section
\ref{sec:scalarfielddynamics}, we define the model, write the
scalar-field equations of motion, show that the dynamics are
those of a nonlinear-sigma model, and introduce the large-$N$
scaling limit for the nonlinear sigma model.  In Section
\ref{sec:matterperturbations}, we write the relation between the
matter-density perturbation and the scalar-field perturbation.
In Section \ref{sec:powerspectrum}, we derive the power spectrum
for density and curvature perturbations, discuss the
normalization, and derive current constraints to the $v$-$N$
parameter space from upper limits to the SOSF contribution to
CMB fluctuations.  In Section \ref{sec:bispectrum}, we discuss the
calculation of the bispectrum, the central focus of this paper.
We present a simplified version, our Eq.~(\ref{eqn:biresult}), of the
matter-bispectrum expression in
Ref.~\cite{Jaffe:1993tt}, evaluate it numerically, and provide a
simple analytic approximation for the results.  We write the
bispectra for matter and curvature perturbations, define a
non-Gaussianity parameter $\fnl^\sigma$ for the model, and
estimate the current constraint to $\fnl^\sigma$ from the CMB.
Section \ref{sec:galaxysurveys} presents the matter bispectrum
for modes that entered the horizon during radiation domination,
those relevant for galaxy surveys.
The central results of the paper are
Eq.~(\ref{eqn:centralresult}) for the curvature bispectrum;
Eq.~(\ref{eqn:fnlsigma}) which defines $\fnl^\sigma$ in terms of
the SOSF model parameters $v$ and $N$;
Eq.~(\ref{eqn:g3approx}) which approximates
the bispectrum function $g_3(k_1,k_2,k_3)$; and
Eqs.~(\ref{eqn:galaxybispectrum}) and
(\ref{eqn:galaxybispectrum2}) which present the matter
bispectrum in a form useful for galaxy surveys.  We make
concluding remarks in Section \ref{sec:discussion}.  An Appendix
contains some calculational details and useful approximations.

\section{Scalar-Field Dynamics}
\label{sec:scalarfielddynamics}

Self-ordering scalar fields are described by an $N$-component
scalar field with an $O(N)$ symmetry that is spontaneously
broken to $O(N-1)$.\footnote{We assume that the issues about
global symmetries raised in Refs.~\protect\cite{Kamionkowski:1992mf}
are somehow solved \protect\cite{Barr:1992qq}.}  After symmetry
breaking, the scalar field
lies in different places in its $S^{N-1}$ vacuum manifold
in different causally disconnected regions of the Universe.  As the Universe
expands and these previously causally-disconnected regions come
into causal contact, field gradients tend to align the scalar
field.  The rate of alignment for these fields is limited only by causality, and
so the fields become aligned within a few Hubble times after
horizon crossing.  Still, as the Universe expands, there are
continually new causally-disconnected regions, on ever larger
scales, that enter the horizon.  The result is thus 
a continual scale-invariant generation of new scalar-field
perturbations.  In this Section, we describe the scalar-field
dynamics; the following Section then describes how the
gradient energy density in these scalar fields induce
perturbations to the matter density.

The starting point is an $N$-component scalar field
$\vec\Phi=(\phi^1,\phi^2,\cdots,\phi^N)$, with $\phi^a$ real, with
Lagrangian density,
\begin{equation}
    \LL =  -(\nabla_\mu\vec\Phi)\cdot (\nabla^\mu\vec\Phi)
    - \frac{\lambda}{4}\left(|\vec\Phi|^2-v^2\right)^2,
\label{eqn:Lagragian}
\end{equation}
where $\lambda$ is the dimensionless self-coupling of $\vec\Phi$,
and $v$ is the magnitude of the vacuum expectation value (vev)
in the true vacuum.
At temperatures $T\ll \lambda^{1/4} v$, the $O(N)$ symmetry of
the Lagrangian is spontaneously broken, and the field is
thereafter restricted to the $S^{N-1}$ vacuum manifold.  The dynamics is
thus effectively that of $N-1$ massless Nambu-Goldstone modes
which we describe in terms of the $N$ fields $\phi^a$ with
the effective Lagrangian density,
\begin{equation}
    \LL =  -(\nabla_\mu\vec \Phi)\cdot(\nabla^\mu\vec\Phi) +
    \Lambda(|\vec\Phi|^2-v^2),
\label{eqn:effectiveLagrangian}
\end{equation}
where $\Lambda$ is a Lagrange multiplier that enforces the
constraint $|\vec\Phi|^2 = v^2$.  The resulting equations of
motion are
\begin{eqnarray}\label{eq:NLSM}
    {\phi^a}''({\mathbf x},\eta) &+&
    2\mathcal{H}{\phi^a}'({\mathbf x},\eta) \nn \\
    &   - & \left[\nabla^2 +
    \frac{1}{v^2}
    (\nabla_\mu\vec\Phi)\cdot(\nabla^\mu\vec\Phi)
    \right]\phi^a({\mathbf x},\eta) = 0,\nn \\
\end{eqnarray}
where the primes denote derivatives with respect to conformal
time $\eta$, and $\mathcal{H}=a'/a$ in terms of the
Friedmann-Robertson-Walker scale factor $a(\eta)$.  Also,
$\nabla^2$ is here a spatial Laplacian in comoving coordinates.
Eq.~(\ref{eq:NLSM}) represents the non-linear sigma model (NLSM
from now on), that describes the evolution of the scalar field
after spontaneous symmetry breaking. 

In the large-$N$ limit, the field components become independent of
each other (up to corrections of order $N^{-1}$).  We thus
replace the bilinear term in the equation of motion by an
ensemble average,
\begin{equation}
    (\nabla_\mu\vec\Phi)\cdot(\nabla^\mu\vec\Phi) = 
    N \VEV{
    (\nabla_\mu\phi^a)(\nabla^\mu\phi^a) }
    \equiv T(\eta)\,,
\label{eqn:scalinganzatz}
\end{equation}
where there is no sum on $a$ in the second equality, and in the
last equality we have made the usual ergodic assumption,
replacing the ensemble average by a spatial average.

The only timescale in the problem is that set by the (comoving)
horizon $\mathcal{H}^{-1} \propto \eta$, so by dimensional
considerations $T \propto \mathcal{H}^2$, and $T(\eta) =
T_o/\eta^{2}$,
with $T_o>0$. We then replace the non-linear term in the
NLSM equation of motion, Eq.~(\ref{eq:NLSM}), by this expectation
value and in this way linearize the equations of
motion. Introducing $\alpha = d\log
a/d\log\eta$ and Fourier transforming the spatial dependence of
the equations, 
\begin{equation}
\phi^a(\bk,\eta) = \int d^3x\, \phi^a(\mathbf{x},\eta)
e^{+i\bk\cdot\bx}\, ,
\end{equation}
we obtain
\begin{eqnarray}\label{e:sigma2}
    \phi_k^{a\,''} + \frac{2\alpha}{\eta}\phi_k^{a\,'}
      + \left(k^2-\frac{T_o}{v^2\eta^2}\right)\phi_k^a = 0\,,
\end{eqnarray}
with $\alpha=1$ for a radiation-dominated Universe and
$\alpha=2$ for a matter-dominated Universe.
For constant $\alpha$, the solution to Eq.~(\ref{e:sigma2})
that is finite as $\eta\to0$, is $\phi^a(\bk,\eta) =
v \epsilon^a(\bk) f(k\eta)$, with $f(x)\equiv x^{1/2-\alpha} J_\nu(x)$,
and $J_\nu(x)$ is a Bessel function.  Here, $\epsilon^a(\bk)$ is the
amplitude of mode $\bk$, and $\nu$ is fixed by $\nu^2 =
(1/2-\alpha)^2+(T_o/v^2)$.

In the large-$N$ limit, the statistical distribution of each
field component approaches a Gaussian distribution with
mean $\VEV{ \phi^a(\bx,\eta)} = 0$ and
variance $\VEV{ \phi^a(\bx,\eta)\phi^b(\bx,\eta)} =
(v^2/N)\delta_{ab}$. The initial field component
$\phi^a(\bx,\eta=0)$
takes on a random value at each point in space.  We thus take
the $\lbrace \epsilon^a(\bk)
\rbrace$ to be Gaussian random variables with mean
$\VEV{\epsilon^a(\bk)}=0$ and variance,
\begin{equation}\label{eq:correlator_a}
     \VEV{ \epsilon^a(\mathbf{k})\epsilon^b(\mathbf{k'}) }
     =
     (2\pi)^3|\mathbf{k}|^{-n}\frac{\delta_{ab}}{A N}\delta_D(\mathbf{k}
     + \mathbf{k'})\, ,
\end{equation}
where $\delta_D(\bk)$ is the Dirac delta function, and $A$ is a
normalization constant to be determined below.  The power-law
dependence on $k$ is taken since the initial conditions are
scale-free.

The power-law index $n$ in Eq.~(\ref{eq:correlator_a}) is fixed
by the condition that $\VEV{  \phi^a(\bx,\eta)\phi^b(\bx,\eta)}
= (v^2/N)\delta_{ab}$ for all $\eta$:
\begin{eqnarray}
    \VEV{ \phi^a(\bx,\eta) \phi^b(\bx,\eta)} &=&
    \delta_{ab} v^2
    \int\frac{d^3k}{(2\pi)^3} \int\frac{d^3k'}{(2\pi)^3}\nn \\
    &  \times& 
    \VEV{ \epsilon^a(\mathbf{k})\epsilon^a(\mathbf{k'})}
    f(k\eta)f(k'\eta) \nn \\
    &=& \frac{4\pi v^2 \delta_{ab}}{(2\pi)^3 AN} \int\,
    dk\, k^{2-n} \,f^2(k\eta).
\label{eq:beta2}  
\end{eqnarray}
We see that $n=3$ gives a result that is independent of time,
and so we choose $n=3$ hereafter.

Just after symmetry breaking, at conformal time $\eta_*$, the
field correlation is then
\begin{eqnarray}
    \VEV{
    \phi^a(\mathbf{k},\eta_*)\phi^b(\mathbf{k'},\eta_*)}
    & \propto & f^2(x_*)\VEV{
    \epsilon^a(\mathbf{k})\epsilon^b(\mathbf{k'})} \nn \\
    &\propto&
    \eta_*^3 |\mathbf{k}|^{1-2\alpha+2\nu-3}\delta_D(\mathbf{k}
    + \mathbf{k'})\, .\nn \\
\label{eqn:powerspectrum}
\end{eqnarray}
Since the initial values $\phi^a(\bx,\eta_*)$ are uncorrelated
on scales $k\ll \eta_*^{-1}$, we set $\nu = \alpha + 1$,
so that the initial field is described by a white-noise power
spectrum.  This then fixes $(T_o/v^2) = 3\alpha + (3/4)$.

We now return to Eq.~(\ref{eq:beta2}) to fix the normalization
constant $A$.  From
\begin{eqnarray}
    \VEV{ \phi^a(\bx,\eta)\phi^b(\bx,\eta)} &=& \frac{v^2
    \delta_{ab}}{2\pi^2
    AN}\int_0^\infty dx\, x^{-2\alpha} J_{\alpha+1}^2(x) \cr
    &=&
    \frac{v^2 \delta_{ab} }{N},
\label{eq:beta3}
\end{eqnarray}
we find
\begin{equation}
    A = \frac{1}{8\pi^2}
    \frac{\Gamma(\alpha)}{\Gamma(2\alpha+3/2)\Gamma(\alpha+1/2)}\,.
\end{equation}
For $\alpha=2$ (matter domination), $A= 16/2835 \pi^3 = 1.82\times10^{-4}$, and
for $\alpha=1$ (radiation domination), $A= 2/15 \pi^3 = 4.3\times10^{-3}$.

\section{Matter-Density Perturbations}
\label{sec:matterperturbations}

Although the scalar field will initially take on different
values in different causally-disconnected regions, the curvature
perturbation is initially zero.  The scalar-field
gradient-energy perturbation that arises as previously
causally-disconnected regions come into causal contact
is then compensated by a perturbation in the matter
density~\cite{Veeraraghavan:1990yd,Jaffe:1993tt}.  

In this Section, we calculate the time evolution of the matter
perturbation.   The action of the scalar field
occurs primarily within a few Hubble times after a particular
Fourier mode $k$ enters the horizon.  The subsequent evolution
of the mode is then governed by gravitational infall as if it were a
primordial perturbation; i.e., the perturbation amplitude
grows only logarithmically during radiation
domination, and then grows with the scale factor during matter
domination.  Our strategy here will be to evaluate
the matter-perturbation amplitude several Hubble times after
horizon crossing, a calculation that is relatively
straightforward.  Strictly speaking, our calculation applies
only to modes that enter the horizon during matter domination, 
but we argue below that our ultimate results for the bispectrum
should also be roughly valid for the smaller-scale modes that
enter the horizon during radiation domination, those relevant
for galaxy surveys.

As described in Ref.~\cite{Jaffe:1993tt}, the scalar-field
alignment involves density perturbations that then lead to
gravitational-potential perturbations which in turn induce the
perturbations to the matter density that are our ultimate
interest.  Following Ref.~\cite{Jaffe:1993tt}, the
matter-density perturbation induced by the scalar field for
modes that enter the horizon during matter domination is
\begin{equation}
    \delta(\bx,\eta) =
    \frac{2\pi G}{5}\eta^2\int
    d\eta' \,\partial_iT_{0i}(\bx,\eta')\,, 
\label{eqn:matterperturbation}
\end{equation}
where $G$ is Newton's constant, and
\begin{equation}
    T_{0i} = (\partial_0{\phi^a})(\partial_i\phi^a)\,,
\end{equation} 
is the $0i$-component of the stress-energy tensor of the
multicomponent scalar field.  The integral in
Eq.~(\ref{eqn:matterperturbation}) approaches a constant for
$\eta \gg \mathrm{few}/k$---i.e., within a few Hubble times after
horizon crossing.  The subsequent $\eta^2$ evolution in the
prefactor is then simply the $\delta \propto a \propto \eta^2$
linear-theory growth of the perturbation amplitude in a
matter-dominated Universe.

Using $G=1/\Mpl^2$, where $\Mpl=1.22\times10^{19}$~GeV is the
Planck mass, and defining
\begin{equation}
    C \equiv \frac{2\pi}{5}\left(\frac{v}{\Mpl}\right)^2\, ,
\end{equation}
the Fourier transform of the density perturbation is
\begin{eqnarray}
    \delta(\bk,\eta)  &=& -{C\eta^2}\int
    \frac{d^3q}{(2\pi)^3}\epsilon^a(\bq)
    \epsilon^a(\bk-\bq)\,|\bk-\bq|  \, (\bk\cdot\bq)\,\nn \\
    & & \times \int
    d\tau\, f'(|\bk-\bq|\tau)f(q\tau)\,, 
\label{eqn:delta_k}
\end{eqnarray}
where $f'(y)\equiv df/dy$. The crucial qualitative feature is
that $\delta(\bk,\eta)$ is quadratic in powers of
$\epsilon^a(\bk)$.  And since $\epsilon^a(\bk)$ is a nearly Gaussian
field, the density field $\delta(\bx)$ will be highly non-Gaussian.

\section{The Power Spectrum}
\label{sec:powerspectrum}

The power spectrum $P^\sigma(k)$ for matter-density
perturbations induced by the scalar field is defined by
\begin{equation}
    \VEV{ \delta(\bk)\delta(\bk')} = (2\pi)^3\,
    \delta_D(\bk+\bk')\, P^\sigma(k),
\label{eqn:PSdefinition}
\end{equation}
where the angle brackets denote an average over all realizations
of the random field $\delta(\bk)$.  The calculation of the power
spectrum is lengthy but straightforward; details are provided in
the Appendix.  The result, given in Eq.~(\ref{eqn:PSresult}), can
be re-written,
\begin{eqnarray}\label{eq:prelimPowerSpectrum}
    P^\sigma(k,\eta) &\equiv&
    \frac{C^2\eta^4}{(2\pi)^2A^2}\frac{k}{N}\int_0^\infty 
    dv\,v^3\nn \\
    &  \times & \int_{-1}^1\,
    dl\,\mathcal{I}(v,b)\,l\,\left[\mathcal{I}(v,b)\,v\,l +
    \mathcal{I}(b,v)(1-v\,l)\right]\, ,\nn \\
\end{eqnarray}
where $b = \sqrt{1+v^2-2v\,l}$, and
\begin{equation}
    \mathcal{I}(a,b) \equiv \int ds
    \frac{f(as)f'(bs)}{a^{3/2}b^{1/2}}\,.
\label{eqn:mathcalI}
\end{equation}
Strictly speaking, the upper limit in this integral is $k\eta$.
However, here we will restrict our attention to modes that have
evolved well within the horizon, $k\eta \gg1$, and so we
take the upper limit of the integral in Eq.~(\ref{eqn:mathcalI})
to be infinity.  In this case, the integral $I(a,b)$ is
antisymmetric in its arguments, and the power spectrum can be
written
\begin{equation}\label{eq:finalPowerSpectrum}
    P^\sigma(k,\eta) \equiv
    \frac{C^2 \eta^4}{A^2}\frac{k}{N} g_2,
\end{equation}
where
\begin{equation}
    g_2 \equiv \int \frac{
    d^3v}{(2\pi)^3}\,\left[\mathcal{I}(v,|\hat\bz-\bv|)
    \right]^2\,(\hat\bz\cdot\bv)\,\left[2(\hat\bz\cdot\bv)-1
    \right]\,,  
\end{equation}
and $\hat\bz$ is a unit vector.
Details on the evaluation of $\mathcal{I}$ are given in the
Appendix.  For $\alpha=2$ (matter domination), the integral
evaluates to $g_2=3.3\times10^{-7}$ and for $\alpha=1$
(radiation domination) it is $g_2=2.1\times10^{-4}$.  Note that
the ratios $g_2/A^2$ that appear in
Eq.~(\ref{eq:finalPowerSpectrum}) are approximately 10 and 11, respectively,
for $\alpha=2,\,1$, implying that the amplitude of the
matter perturbation induced by the unwinding of the scalar field
is the same, to $O(10\%)$, for modes that enter the horizon during
matter and radiation domination.

\subsection{Normalization of the power spectrum}
\label{sec:normalization}

We now estimate the constraints to the $v$-$N$ parameter space
from the empirical constraint that the SOSF provide no more than
a fraction $p_\sigma\simeq0.1$ to $C_{l=10}$ \cite{Bevis:2004wk}, the
CMB temperature power spectrum at multipole moment $l=10$.  

On subhorizon scales during matter domination, the curvature
perturbation $\zeta(\bx)$ is related to the gravitational
potential $\Phi(\bx)$ by $\zeta(\bx)=(5/3)\Phi(\bx)$.  The
gravitational potential is related to the density perturbation
through the Poisson equation, $\nabla^2 \Phi = 4\pi G a^2 \bar\rho
\delta$, where $\bar\rho$ is the mean density.  In Fourier
space, the curvature perturbation $\zeta(\bk)$ is thus related
to the matter-density perturbation $\delta(\bk)$ by
\begin{equation}
    \zeta(\bk) = -\frac{5}{2} \left( \frac{aH}{k} \right)^2
    \delta(\bk)\,,
\end{equation}
where we have used the Friedmann equation $H^2 = 8\pi G \rho/3$,
and $H=(da/dt)/a$ is the expansion rate.  The amplitude of the
curvature power spectrum due to the SOSF is therefore,
\begin{eqnarray}
    \Delta_{\mathcal{R}\sigma}^2 &\equiv& \frac{k^3}{2\pi^2}
    P_\zeta(k) = \frac{k^3}{2\pi^2} \left[\frac{5}{2}
    \left(\frac{ a H}{k} \right)^2 \right]^2 P(k) \nn \\
    &=& 8 \left( \frac{v}{\Mpl} \right)^4 \frac{g_2}{ A^2 N}
    \simeq 80\, \left( \frac{v}{\Mpl} \right)^4
    \frac{1}{N}\, ,
\end{eqnarray}
where we have used $(aH\eta)=2$ during matter domination.
The next step is then to determine the relation
between the curvature--power-spectrum amplitude
$\Delta_{\mathcal{R}\sigma}^2$ and the temperature-fluctuation
amplitude.  This is a notoriously difficult calculation, but to
get an estimate, we use Fig.~4 in Ref.~\cite{Pen:1997ae}, which shows
that the large-angle temperature fluctuation $\Delta T$ in a SOSF
model is $G_{\mathrm{sw}}\simeq10$ times greater than it would be in an
adiabatic model with the same matter--power-spectrum
normalization on large scales.\footnote{The factor of 10 is a bit
larger than the factor of 6 one might attribute due to the
difference (1/3 versus 2) for the Sachs-Wolfe amplitude for
adiabatic and isocurvature perturbations.  The additional
$\Delta T$ may be due in part to the vector and tensor
perturbations that are also excited in SOSF models.}  Current
CMB measurements indicate a curvature power spectrum
$\Delta_{\mathcal{R}}\simeq 5\times10^{-5}$, if primordial
perturbations are adiabatic.  If the SOSF provides $G_{\mathrm{sw}}$
times more $\Delta T$ for fixed curvature, and if they make a
fractional contribution $p_\sigma$ to the large-angle
temperature variance, then $\Delta_{\mathcal{R}\sigma}^2 =
(p_\sigma/G_{\mathrm{sw}}^2) \Delta_{\mathcal{R}}^2$.  We thus obtain
\begin{equation}
    \frac{v}{N^{1/4}} = \left(\frac{p_\sigma A^2
    \Delta_{\mathcal{R}}^2} { 8 G_{\mathrm{sw}}^2
    g_2} \right)^{1/4} \Mpl
    \lesssim \frac{\Mpl}{2000}\,,
\label{eqn:vnormalization}
\end{equation}
where the numerical result is obtained by taking $p_\sigma=0.1$
and $G_{\mathrm{sw}}=10$.  The numerical upper limit in
Eq.~(\ref{eqn:vnormalization}) is in good agreement with limits
obtained from simulations \cite{Pen:1993nx}.

\section{The Bispectrum}
\label{sec:bispectrum}

\subsection{The calculation}
\label{sec:thecalculation}

The calculation of the bispectrum proceeds analogously.  The
matter bispectrum $B(k_1,k_2,k_3)$ is defined by
\begin{equation}
   \VEV{\delta(\bk_1)\delta(\bk_2)\delta(\bk_3)} = (2\pi)^3
   \delta_D(\bk_1 + \bk_2+\bk_3) B(k_1,k_2,k_3).
\label{eqn:bispectrumdefinition}
\end{equation}
Although the definition of the bispectrum is nominally in terms
of the vector quantities $\bk_i$, the triangle constraint
$\bk_1+\bk_2+\bk_3=0$ imposed by the Dirac delta function, as well
as statistical isotropy, imply that the bispectrum is most
generally a function of the magnitudes $k_i$ of the three sides
of the triangle.  Again, some details of the calculation are
provided in the Appendix.  The result is
\begin{equation}
    B(k_1,k_2,k_3) = \frac{C^3 \eta^6}{A^3 N^2}
    g_3(k_1,k_2,k_3)\,,
\label{eqn:ourbispectrum}
\end{equation}
where
\begin{eqnarray}
    g_3(k_1,k_2,k_3) &\equiv& \int
    \frac{d^3v}{(2\pi)^3}  H(\bu+\bv,\bv) \nn \\
    & & \times  H(\bv,\hat\bz-\bv) H(\hat\bz-\bv,\bu+\bv),
\label{eqn:biresult}
\end{eqnarray}
with
\begin{equation}
    H(\mathbf{a},\mathbf{b}) \equiv \mathcal{I}(a,b)(b^2-a^2)\,.
\end{equation}
We have chosen $\vec k_1$ in
Eq.~(\ref{eqn:biresult}) to be in the $\hat\bz$
direction, without loss of generality, and we have then defined
$\bu\equiv \bk_2/k_1$.  Note that $H(\vec a,\vec
b)=H(a,b)=H(b,a)$; i.e., it is a function only of the magnitudes
of its arguments, and it is symmetric in its arguments.  Note
further that $H(a,b)\leq0$, and thus $g_3(k_1,k_2,k_3)<0$.
The function $g_3(k_1,k_2,k_3)$ depends
only on the shape of the triangle, not on its overall
size---i.e., $g_3(k_1,k_2,k_3)=g_3(1,k_2/k_1,k_3/k_1)$---a
consequence of the scale-invariance of SOSFs.  We have checked
that Eq.~(\ref{eqn:biresult}) is equivalent to, although far
simpler, than Eq.~(59) in Ref.~\cite{Jaffe:1993tt}.  Given the
symmetry of $H(a,b)$ in its arguments, it is simple to check
that $g_3(\bw -\hat\bz-\bu)=g(\bu)$, as it should
(given that the three sides of the triangle should add as $\hat
\bz + \bu +\bw=0$).  If we set the third side to have length
$w=k_3/k_1$, then $\cos\theta \equiv \bu\cdot \hat \bz =
(w^2-1-u^2)/(2u)$.  If we choose $k_1\geq k_2 \geq k_3$, then
$\cos\theta < -(2u)^{-1}$.

\begin{figure}[htbp]
\resizebox{!}{5cm}{\includegraphics{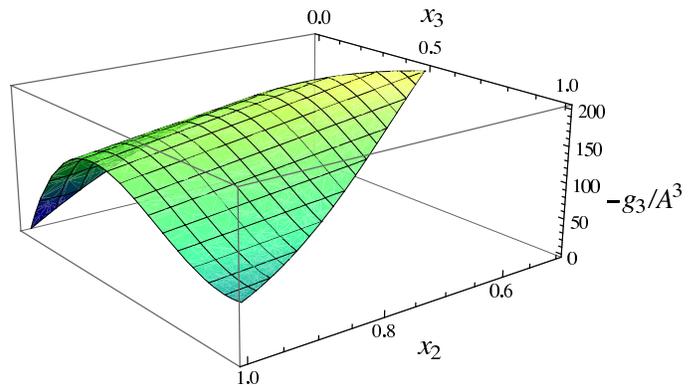}}
\caption{The function $-g_3(k_1,k_2,k_3)$, taking $k_1=1$, for
    modes that enter the horizon during matter domination.  The
    figure looks virtually identical for modes that enter the
    horizon during radiation domination.}
\label{fig:g3}
\end{figure}

We have calculated $g_3(k_1,k_2,k_3)$ numerically, and the
result is shown in Fig.~\ref{fig:g3}.  We note (prefacing the
discussion below) that the quantity, $-g_3(1,x_2,x_3)$, with
$x_2=k_2/k_1$ and $x_3=k_3/k_1$, that we plot is the same (up
to some normalization factor) as the quantity
$F(q,x_2,x_3)x_2^2 x_3^2$ plotted in Figs.~1 and 2 in
Ref.~\cite{Creminelli:2005hu} which show, respectively, the
bispectra for the local-model and equilateral model.  Those
figures show that the local-model bispectrum peaks sharply for
``squeezed'' triangles ($k_1\simeq k_2 \gg k_3$) and that the
equilateral-model bispectrum peaks at equilateral triangles
($k_1\simeq k_2 \simeq k_3$).  Our Fig.~\ref{fig:g3} shows that
the SOSF bispectrum is, however, quite different.  It is nonzero
for equilateral triangles, goes to zero in the squeezed limit,
and it peaks for ``aligned'' triangles, $k_1 \simeq 2 k_2 \simeq
2 k_3$.

To aid in data-analysis
efforts, we have found that the following approximation
reproduces the numerical results for $g_3(k_1,k_2,k_3)$ to
within a few percent:
\begin{eqnarray}
    g_3(k_1,k_2,k_3) &=& -\frac{A^3}{143} \left(262 - 127
    \frac{k_2}{k_1} \right) \nn \\
    &\times&\left[  947 \frac{k_3}{k_1}
    - 1770 \left(\frac{k_3}{k_1}\right)^2 + 893 \left(\frac{k_3}{k_1}
    \right)^3\right], \nn \\
\label{eqn:g3approx}
\end{eqnarray}
where we take $k_1 \geq k_2 \geq k_3$ in this expression.

\subsection{Curvature Bispectrum}
\label{sec:curvaturebispectrum}

To compare with results for other models, and for comparison
with CMB constraints, we next calulate the curvature bispectrum
$F(k_1,k_2,k_3)$, defined by
\begin{equation}
    \VEV{\zeta(\bk_1)\zeta(\bk_2)\zeta(\bk_3)} = (2\pi)^3
    \delta_D(\bk_1+\bk_2 +\bk_3) F(k_1,k_2,k_3).
\label{eqn:curvbidefn}
\end{equation}
Using the relations above, we can write $C^2 = (\pi^2/50) (A^2 N
\Delta_{\mathcal{R}\sigma}^2 / g_2)$ and then find,
\begin{equation}
    F^\sigma(k_1,k_2,k_3) = - \frac{2\sqrt{2} \pi^3
    \Delta_{\mathcal{R}\sigma}^3 }{ g_2^{3/2} N^{1/2}} \frac{
    g_3(k_1,k_2,k_3)}{k_1^2 k_2^2 k_3^2}\, .
\label{eqn:almostcentralresult}
\end{equation}
A few observations:  (1) It is only the amplitude, not the
shape, that depends on the symmetry-breaking scale $v$.
(2) The bispectrum decreases, for fixed
$\Delta_{\mathcal{R}\sigma}$, as $N^{-1/2}$ 
with increasing $N$, again reflecting that the model should
become increasingly Gaussian with more fields, a consequence of
the central-limit theorem.  (3) The scaling with $\Delta_{\mathcal{R}}$
is $\propto \Delta_{\mathcal{R}}^3$, as opposed to the
$\Delta_{{\mathcal{R}}}^4$ scaling of the local-model bispectrum.  In
words, the non-Gaussianity is of order unity, a consequence of
the fact that the density perturbation is the square of a Gaussian
random field [cf., Eq.~(\ref{eqn:matterperturbation})], rather
than something very small, as in inflationary models.

We now put the curvature bispectrum in a slightly more familiar
form by defining a non-Gaussianity parameter $\fnl^\sigma$ for
the model.  To do so, we recall that the local-model prediction
for the curvature bispectrum is,
\begin{eqnarray}
   F^{\mathrm{local}}(k_1,k_2,k_3) &=& 2\frac{3}{5} (2\pi^2)^2
   \Delta_{\mathcal{R}}^4 \fnl^{\mathrm{local}} \nn \\
   & & \times \left[ \frac{1}{k_1^3 k_2^3}
   +\frac{1}{k_2^3 k_3^3} +\frac{1}{k_1^3 k_3^3}  \right]\, ,
\label{eqn:localmodelcurvbi}
\end{eqnarray}
where $\fnl^{\mathrm{local}}$ is the local-model non-Gaussianity
parameter, defined by writing the curvature as $\zeta(\bx) =
\zeta_g(\bx) + (3/5) \fnl^{\mathrm{local}} \left[
(\zeta_g)^2(\bx)-\VEV{ (\zeta_g)^2(\bx)} \right]$ in terms of a
Gaussian field $\zeta_g(\bx)$.  

We now define the non-Gaussianity parameter $\fnl^\sigma$ for
SOSFs by equating the local-model and SOSF bispectra for
equilateral triangles; i.e., equating
Eqs.~(\ref{eqn:localmodelcurvbi}) and
(\ref{eqn:almostcentralresult}), we define,
\begin{eqnarray}
    \fnl^\sigma &\equiv&  -\frac{5\, p_\sigma^{3/2} g_3(1,1,1)}{18
    \sqrt{2}\,\pi\, N^{1/2}
    g_2^{3/2} \Delta_{\mathcal{R}} G_{\mathrm{sw}}^3} \nn \\
    &\simeq&  40\,G_{\mathrm{sw}}^{-3} \left(\frac{
    p_\sigma}{0.1} \right)^{3/2}
    \left(\frac{N}{5}\right)^{-1/2} \nn \\
    & \simeq& 3\, G_{\mathrm{sw}}^{-3} \left(\frac{v}{5\times10^{15}\,
    \mathrm{GeV}} \right)^6 \left(\frac{N}{5} \right)^{-2}\,.
\label{eqn:fnlsigma}
\end{eqnarray}
With this $\fnl^\sigma$, the curvature bispectrum can then be
written
\begin{equation}
    F(k_1,k_2,k_3) = \frac{18}{5} (2\pi^2)^2
    \Delta_{\mathcal{R}}^4 \frac{ \fnl^\sigma}{k_1^2 k_2^2
    k_3^2} \frac{ g_3(k_1,k_2,k_3)}{ g_3(1,1,1)}\,.
\label{eqn:centralresult}
\end{equation}
Similarly, the gravitational-potential bispectrum can be
obtained by multiplying this expression by $5/3$ and then
replacing $\Delta_{\mathcal{R}}^2$ by $\Delta_{\Phi}^2=(3/5)^2
\Delta_{\mathcal{R}}^2$.  Note that $\fnl^\sigma$ is manifestly
positive, unlike $\fnl$ for the local or equilateral models,
which may take on either sign.

\subsection{Estimate of CMB Constraints}
\label{sec:currentconstraints}

As indicated in the Introduction, the bispectrum can be probed
with the CMB, large-scale structure, and the abundances of
objects.  The strongest current constraints to the local-model
bispectrum come from the CMB \cite{limits}, followed closely by
galaxy-clustering constraints.  Given that SOSFs produce a
larger temperature-fluctuation amplitude for a given
density-perturbation amplitude, we surmise that the CMB will
provide stronger constraints to SOSF non-Gaussianity than galaxy
clustering.  We thus now estimate a constraint to $\fnl^\sigma$
from the CMB.

Before doing so, we first caution that Eq.~(\ref{eqn:centralresult})
is derived for the curvature perturbation only for modes once
they are well within the horizon.  It is thus not, strictly
speaking, appropriate for CMB modes $l\lesssim 100$.  Still, the
shape dependence of the bispectrum, and its amplitude relative
to the curvature-perturbation amplitude, arises primarily from
the quadratic dependence of the density perturbation on the
scalar-field perturbation as encoded in
Eq.~(\ref{eqn:delta_k}).  The shape dependence of the bispectrum
we calculate should thus be at least roughly correct even for
$l\lesssim100$.

CMB constraints to $\fnl$ are typically applied assuming that
the density perturbations are adiabatic, which implies a certain
relation, $(\Delta T/T) \simeq -\zeta/5$, for the large-angle
temperature fluctuation.  In our case, though, there is roughly
$G_{\mathrm{sw}}\simeq 10$ times more $\Delta T$ for a given $\zeta$
than in adiabatic models.  If so, and if all of these
temperature fluctuations are due to scalar perturbations (rather
than vector and/or tensor modes), then the implied CMB
bispectrum should be roughly $G_{sw}^3$ times larger.
Simulations show, though, that only a fraction $f_{\mathrm{s}}\simeq0.5$ of
the SOSF temperature-fluctuation power is due to scalar modes,
the rest coming from vector and tensor perturbations  \cite{Pen:1997ae,Turok:1997gj}.  The
implied CMB bispectrum should thus scale with $f_{\mathrm{s}}^{3/2}$.
Combining these scalings with $\fnl^\sigma\propto
G_{\mathrm{sw}}^{-3}$ [see Eq.~(\ref{eqn:fnlsigma})], the
$G_{\mathrm{sw}}$ dependence of the CMB bispectrum drops out.
We can therefore apply CMB constraints to $\fnl^\sigma$ by
identifying the $\fnl$ constraints obtained from the CMB for
adiabatic perturbations with $G_{\mathrm{sw}}^3 f_{\mathrm{s}}^{3/2}
\fnl^\sigma$.  And one final caveat: We disregard the
differences in the temperature power spectra from SOSFs and
adiabatic perturbations.

Keeping these multiple caveats in mind, we proceed with our very
rough estimate by noting that the WMAP-7 95\% C.L.\ limit to
$\fnl^{\mathrm{equil}}$, the non-Gaussianity parameter for the
equilateral model, is $ -211 < \fnl^{\mathrm{equil}} < 266$
\cite{Komatsu:2010fb};
this bispectrum is maximized for equilateral triangles.  On the
other hand, the SOSF bispectrum is maximized for aligned 
triangles and is zero for squeezed triangles.  We thus 
conclude that the constraint to $G_{\mathrm{sw}}^3 f_{\mathrm{s}}^{3/2} \fnl^\sigma$
will be stronger than that to $\fnl^{\mathrm{equil}}$, but it is
not clear---given the different weightings to squeezed and
aligned triangles---how it will compare with that to
$\fnl^{\mathrm{local}}$.  Applying these rough arguments to Eq.~(\ref{eqn:fnlsigma}), 
with $f_{\mathrm{s}} \simeq 0.5$, then we
estimate a non-Gaussianity parameter in excess of the predicted threshold $\fnl \sim 7$
for detection by Planck \cite{Cooray:2008xz}. For now, however, we simply estimate
conservatively that $| G_{\mathrm{sw}}^3 f_{\mathrm{s}}^{3/2} \fnl^\sigma | \lesssim
200$. 

\section{The Bispectrum for Galaxy Surveys}
\label{sec:galaxysurveys}

We have carried out our calculations in the regime where
analytic progress is most easily made---i.e., modes that have
entered the horizon during matter domination and only after
those modes have evolved well within the horizon.  Strictly
speaking, therefore, our calculations apply only to galaxy
surveys on {\it very} large scales---those generally larger than
extant surveys cover---and possibly to mid-scale regimes in the
CMB.  

Still, our results for the bispectrum can be easily adapted to
obtain roughly the bispectrum for smaller-scale modes, those
that enter the horizon during radiation domination and those
relevant for galaxy surveys.  The calculation for the evolution
of these modes is altered by three effects: (1) The index
$\nu=1+\alpha$ for the Bessel functions in the scalar-field
dynamics (Section \ref{sec:scalarfielddynamics}) is different.
However, we have shown that this has no more than an $O(10\%)$
effect on the power-spectrum and bispectrum.  This is simply
because the correlations between different Fourier modes of the
scalar-field energy density are imprinted, through
Eq.~(\ref{eqn:delta_k}), by the dependence of those Fourier
modes on the scalar-field perturbations.  This is the same for
modes that enter the horizon during matter and radiation
domination.  The other two effects are (2) a slightly different
amplitude for the matter perturbation, relative to the
scalar-field energy-density perturbation, for modes that enter
the horizon during radiation domination [Eq.~(32) for RD in
Ref.~\cite{Jaffe:1993tt}, as opposed to Eq.~(29) in the same reference, our
Eq.~(\ref{eqn:matterperturbation}), for MD]; and (3) the usual
linear-theory growth of primordial isocurvature perturbations
through radiation domination and the transition to matter
domination.  These latter two effects amount to a calculation of
the transfer function $T(k)$ for the matter 
power spectrum in SOSF models, which can be accomplished either
with simulations [cf., Ref.~\cite{Pen:1997ae,Pen:1993nx}] or
approximately with standard linear-theory calculations with
primordial isocurvature fluctuations.  Again, however, although
the calculation of the evolution of the amplitudes of the
small-scale density-perturbation Fourier modes will be far more
complicated than the larger-scale modes we have focused upon,
the {\it correlations} between those modes will be, at the
$O(10\%)$ level, the same as those we have calculated for
larger-scale modes.

More precisely, all we need to do is replace the density fields
$\delta(\bk)$ in Sections \ref{sec:powerspectrum} and
\ref{sec:thecalculation} by $\delta(\bk)T(k)$, where $T(k)$ is
the SOSF transfer function.  The matter power spectrum
$P^\sigma(k)$ due to SOSFs is then obtained from that in
Eq.~(\ref{eq:finalPowerSpectrum}) by multiplying by $|T(k)|^2$,
and the matter bispectrum is obtained by multiplying that in
Eq.~(\ref{eqn:ourbispectrum}) by $T(k_1)T(k_2)T(k_3)$.  We can
then write the normalization constant $(C \eta^2/A)^3$ in
Eq.~(\ref{eqn:ourbispectrum})  in terms of the (processed)
matter power spectra using Eq.~(\ref{eq:finalPowerSpectrum}) to
obtain the matter bispectrum,
\begin{equation}
    B(k_1,k_2,k_3) = \frac{g_3(k_1,k_2,k_3)}{g_2^{3/2} N^{1/2}}  
    \left[ \frac{ P^\sigma(k_1)P^\sigma(k_2)P^\sigma(k_3)}{ k_1
    k_2 k_3} \right]^{1/2}.
\label{eqn:galaxybispectrum}
\end{equation}
valid for galaxy-survey scales.  Here $P^\sigma(k)$ is the {\it
processed} power spectrum due to SOSFs; i.e., it includes the
transfer function.  Using Eq.~(\ref{eqn:fnlsigma}), this can be
re-written in terms of $\fnl^\sigma$ as,
\begin{widetext}
\begin{eqnarray}
    B(k_1,k_2,k_3)  & = &
\frac{18 \sqrt{2} \pi \fnl^\sigma
    \Delta_{\mathcal{R}} G_{\mathrm{sw}}^3}{5 p_\sigma^{3/2}}
    \frac{g_3(k_1,k_2,k_3)}{g_3(1,1,1)}  \left[ \frac{ P^\sigma(k_1)P^\sigma(k_2)P^\sigma(k_3)}{ k_1
    k_2 k_3} \right]^{1/2} \nn \\
    & \simeq & 
    25\, \fnl^\sigma \left( \frac{G_{\mathrm{sw}}}{10}
    \right)^3 \left( \frac{p_\sigma}{0.1} \right)^{-3/2}
    \frac{g_3(k_1,k_2,k_3)}{g_3(1,1,1)}  \left[ \frac{ P^\sigma(k_1)P^\sigma(k_2)P^\sigma(k_3)}{ k_1
    k_2 k_3} \right]^{1/2}.
\label{eqn:galaxybispectrum2}
\end{eqnarray}
\end{widetext}
We leave further evaluation of this bispectrum, as well as
assessment of current constraints, for future work.

\section{Discussion}
\label{sec:discussion}

If some post-inflationary physics involves the spontaneous
breaking of an exact $O(N)$ symmetry with $N>4$, then the
ordering of these scalar fields may provide a secondary
contribution to primordial perturbations.  Current constraints
allow up to $\sim10\%$ of the power in primordial perturbations to be due to
SOSFs.  SOSF models are appealing from the theoretical perspective 
because they are simple,
well-defined, and parametrized only by the symmetry-breaking
scale $v$ and number $N$ of fields.

In this paper we have calculated the matter and curvature
bispectra induced by the ordering of such scalar fields.  Given
that the density perturbation is quadratic in the scalar-field
perturbation, SOSF density perturbations are expected to be
highly non-Gaussian, and if so, measurements of non-Gaussianity
may provide the means to test these models.

Here we have calculated analytically the bispectrum due to SOSFs
and presented results in a way that should be easily accessible
to those doing measurements with the CMB and large-scale
structure.  We find that the triangle-shape dependence of the
bispectrum peaks for aligned triangles, unlike the local-model
bispectrum, which is largest for squeezed triangles, and the
equilateral bispectrum, which is largest for equilateral
triangles.  We have estimated a current upper limit to the
non-Gaussianity parameter $\fnl^\sigma$ for the model and find
that the implied constraints to the $v$-$N$ SOSF parameter space
are competitive with those from the upper limit to CMB
temperature fluctuations.

Finally, we have already argued above, in Section
\ref{sec:currentconstraints}, that the correlation of modes will
be similar for the large-scale modes as they enter the horizon,
those relevant for large-angle CMB fluctuations.  We therefore
believe that rough constraints to the model can be derived from
CMB measurements by assuming that the curvature bispectrum we
calculate is the primordial one.

Clearly, there is room for further numerical work to test our
assumptions and to make our predictions more precise.  In the
meantime, though, we believe that our analytic approximation
captures the essential physics and that our bispectrum can be
used in the meantime as a ``working-horse'' model to derive
constraints, from non-Gaussianity measurements, to this
interesting class of models for secondary contributions to
primordial perturbations.

Finally, we note that the model makes a number of other
predictions.  Given that density perturbations are actively
generated as new modes come within the horizon, vector and
tensor modes will be excited, and these may give rise to
interesting polarization signals \cite{Seljak:1997ii} in the CMB
and perhaps excite B modes \cite{Kamionkowski:1996ks} in the
CMB that might be distinguished from those due to inflation
\cite{Baumann:2009mq}.  There will also be a scale-invariant
spectrum of primordial gravitational waves produced
\cite{Krauss:1991qu} that can be sought in
gravitational-wave observatories.

\acknowledgments

DGF thanks Caltech and CERN for hospitality during the
completion of this work.  DGF acknowledges support from an FPU
Fellowship from the Spanish Ministry of Science, with
ref. AP-2005-1092.  This work was supported at Caltech by DoE
DE-FG03-92-ER40701 and the Gordon and Betty Moore Foundation and
at Dartmouth by NSF AST-0349213.

\section*{Appendix: Calculational Details}

\subsection{The Power Spectrum}
From Eq.~(\ref{eqn:delta_k}),
the two-point correlator in Eq.~(\ref{eqn:PSdefinition}) 
is expressed in terms of the correlator of 
four $\epsilon^a(\bk)$'s. Since $\epsilon^a(\bk)$'s are 
gaussian distributed, we find via the Wick theorem,
\begin{widetext}
\begin{eqnarray}
    \VEV{\delta(\bk)\delta(\bk')} &=&
    {C^2\eta^4} \int \frac{d^3q\,d^3q'}{(2\pi)^6} \left\langle
    \epsilon^a(\bq) \epsilon^a(\bk-\bq) \epsilon^b(\bq')
    \epsilon^b(\bk'-\bq') \right\rangle\,\nn\\ 
    && \times\,|\bk-\bq||\bk'-\bq'|(\bk\cdot\bq) \,
    (\bk'\cdot\bq') \int d\tau\, f'(|\bk-\bq|\tau)f(q\tau) \int
    d\xi\, f'(|\bk'-\bq'|\xi)f(q'\xi) \nn \\
    &=& {C^2\eta^4} \int \frac{d^3q\,d^3q'}{(2\pi)^6} \left[
    \VEV{ \epsilon^a(\bq) \epsilon^a(\bk-\bq) } \VEV{
    \epsilon^b(\bq') \epsilon^b(\bk'-\bq')} \right. \nn \\
    && 
    + \left. \VEV{\epsilon^a(\bq)\epsilon^b(\bq')} \VEV{
    \epsilon^a(\bk-\bq)\epsilon^b(\bk'-\bq')} +
    \VEV{
    \epsilon^a(\bq)\epsilon^b(\bk'-\bq')} \VEV{
    \epsilon^a(\bk-\bq)\epsilon^b(\bq')}\right] \,\nn\\
    &&
    \times\,|\bk-\bq||\bk'-\bq'|\,(\bk\cdot\bq)\,(\bk'\cdot\bq')
    \int d\tau\, f(q\tau)f'(|\bk-\bq|\tau) \int d\xi\,
    f(q'\xi) f'(|\bk'-\bq'|\xi)\,. \nn \\
\label{eqn:beginning}
\end{eqnarray}
Using Eq.~(\ref{eq:correlator_a}), we find
\begin{eqnarray}
    &&
    (\bk\cdot\bq) (\bk'\cdot \bq') \, \VEV{ \epsilon^a(\bq)
    \epsilon^a(\bk-\bq)} \VEV{
    \epsilon^b(\bq')\epsilon^b(\bk'-\bq')} \nn\\ 
    && = (\bk\cdot\bq) (\bk'\cdot \bq')\, \sum_{a,b}
    \frac{(2\pi)^6}{A^2 N^2 q^3
    q'^3}\delta_{aa}\delta_{bb}\delta_D(\bk)
    \delta_D(\bk') = 0,
\end{eqnarray}
\begin{eqnarray}
    && (\bk\cdot\bq) (\bk'\cdot \bq') \,\left\langle\epsilon^a(\bq)
    \epsilon^b(\bq')\right\rangle \left\langle
    \epsilon^a(\bk-\bq)\epsilon^b(\bk'-\bq')\right\rangle \nn\\  
    && =
    \frac{(2\pi)^6}{N A^2 q^3 |\bk-\bq|^3}\,(\bk\cdot\bq)^2 \,
    \delta_D(\bq+\bq') \delta_D(\bk+\bk')\,,
\end{eqnarray}
and
\begin{eqnarray}
    &&
    (\bk\cdot\bq) (\bk'\cdot \bq') \, \left\langle
    \epsilon^a(\bq)\epsilon^b(\bk'-\bq')\right\rangle \left\langle
    \epsilon^a(\bk-\bq)\epsilon^b(\bq')\right\rangle \nn\\ 
    && =
    \frac{(2\pi)^6}{N A^2 q^3 |\bk-\bq|^3}\,(\bk\cdot\bq)
    \left(\bk\cdot(\bk-\bq)
    \right)\,\delta_D(\bq'-\bq+\bk)\delta_D(\bk+\bk')\,. 
\end{eqnarray}
Substituting these expressions into Eq.~(\ref{eqn:beginning}),
we obtain
\begin{eqnarray}
    P^\sigma(k) &=&
    \frac{C^2\eta^4}{A^2 N}\int \frac{d^3q}{(2\pi)^3}
    \left[ \left(\int d\tau\frac{f(q\tau)
    f'(|\bk-\bq|\tau)}{q^{3/2}|\bk-\bq|^{1/2}}\right)^2 (\bk\cdot\bq)^2
    + \right. \nn \\ 
    && \hspace{1.5cm} + \left.\left(\int
    d\tau\frac{f(q\tau) f'(|\bk-\bq|\tau)}{
    q^{3/2}|\bk-\bq|^{1/2}} \right) \left(\int
    d\xi\frac{f(|\bk-\bq|\xi)f'(q\xi)}
    {|\bk-\bq|^{3/2}q^{1/2}}\right)(\bk\cdot \bq) \left(
    \bk\cdot (\bk-\bq) \right)\right]\,\nn\\
    &=& \frac{C^2\eta^4}{A^2}\,\frac{k}{N}\,\int \frac{d^3v}{(2\pi)^3}\,\mathcal{I}(v,|\hat\bk-\bv|)
    \left[(\hat\bk\cdot\bv)^2\,\mathcal{I}(v,|\hat\bk-\bv|)
   + (\hat\bk\cdot\bv)(1-\hat\bk\cdot\bv)\mathcal{I}(|\hat\bk-\bv|,v)
    \right]\,, 
\label{eqn:PSresult}
\end{eqnarray}
where $\mathcal{I}(a,b)$ 
is defined in Eq.~(\ref{eqn:mathcalI}). 
From here we can then introduce $g_2$ in 
Section~\ref{sec:powerspectrum}.

\subsection{The Bispectrum}

The bispectrum is obtained by starting with,
\begin{eqnarray}
    \left\langle\delta(\bk_1)\delta(\bk_2)\delta(\bk_3)\right\rangle &=&
    -{C^3\eta^6} \int \frac{d^3q_1\, d^3q_2 d^3q_3}{(2\pi)^9} \left\langle
    \epsilon^a(\bq_1) \epsilon^a(\bk_1-\bq_1) 
    \epsilon^b(\bq_2) \epsilon^b(\bk_2-\bq_2) 
    \epsilon^c(\bq_3) \epsilon^c(\bk_3-\bq_3) 
    \right\rangle\,\nn\\ 
    && \times\,|\bk_1-\bq_1| |\bk_2-\bq_2| |\bk_3-\bq_3| (\bk_1\cdot\bq_1) \,
    (\bk_2\cdot\bq_2)  (\bk_3\cdot\bq_3) \nn \\
   && \times \int d\tau_1\,
    f(q_1\tau_1)f'(|\bk_1-\bq_1|\tau_1) 
    \int d\tau_2\,
    f(q_2\tau_2)f'(|\bk_2-\bq_2|\tau_2) 
    \int d\tau_3\,
    f(q_3\tau_3)f'(|\bk_3-\bq_3|\tau_3) \,.\nn \\
\label{eqn:bispectrumstart}
\end{eqnarray}
The expectation value of the product of six $\epsilon^a(\bk)$'s
can be expanded with Wick contractions and, after some algebra,
and  using Eq.~(\ref{eq:correlator_a}), we find
\begin{eqnarray}
    \left\langle\delta(\bk_1)\delta(\bk_2)\delta(\bk_3)\right\rangle 
    &=& -(2\pi)^3 \delta_D(\bk_1+\bk_2+\bk_3)
    \frac{C^3\eta^6}{A^3N^2}\,\int\frac{d^3v}{(2\pi)^3}\,\,
    (\hat\bk\cdot{\bf v})\,\,\mathcal{I}(v,b)\, \nn \\
    &&\times\,\,\left\lbrace\big[({\bf u}\cdot{\bf v})\,\,
    \mathcal{I}(v,b_2) - ({\bf u}\cdot{\bf b}_{2})\,\,
    \mathcal{I}(b_{2},v)\,\big]\,
    \big[\,((\hat\bk+{\bf u})\cdot{\bf b})
    \,\,\mathcal{I}(b,b_2)+((\hat\bk+{\bf u})\cdot{\bf b}_2)\,\,
    \mathcal{I}(b_2,b)\,\big]\right.\nn
    \\
    && \hspace{.5cm} \left.+ \big[({\bf u}\cdot{\bf b}\,\,
    \mathcal{I}(b,b_{12}) - ({\bf u}\cdot{\bf b}_{12})\,\,
    \mathcal{I}(b_{12},b)\,\big]\,
    \big[\,((\hat\bk+{\bf u})\cdot\bv)\,\,
    \mathcal{I}(v,b_{12})+((\hat\bk+{\bf u})\cdot{\bf b}_{12})\,\,
    \mathcal{I}(b_{12},v)\,\big]\right\rbrace , \nn \\
\end{eqnarray}
where $k \equiv |\bk_1|$, $\hat\bk \equiv \bk_1/k$, 
${\bf u} \equiv \bk_2/k$, and we 
have defined ${\bf b} \equiv \hat\bk-\bv, {\bf b}_2 \equiv 
{\bf u} + \bv, {\bf b}_{12} \equiv \hat\bk + {\bf u} - \bv$. 
Defining $g_3$ as in Eq.~(\ref{eqn:ourbispectrum}) and performing 
a change of variables $\bv\,\rightarrow\, (\hat\bk - \bv)$ in 
the second term of the above integral, we then find
\begin{eqnarray}
    g_3(k_1,k_2,k_3) &=& 
    \int\frac{d^3v}{(2\pi)^3}\big[(\hat\bk\cdot\bv)\mathcal{I}(v,b)
    + (\hat\bk\cdot{\bf b})\mathcal{I}(b,v)\big]\big[({\bf
    u}\cdot{\bf v})\mathcal{I}(v,b_2) - ({\bf u}\cdot{\bf
    b}_2)\mathcal{I}(b_2,v)\big] \nn \\
    & &  \hspace{0.5cm} \times \big[{\bf b}\cdot(\hat\bk+{\bf
    u})\mathcal{I}(b,b_2) + {\bf b}_2\cdot(\hat\bk+{\bf
    u})\mathcal{I}(b_2,b)\big] \,.\nn\\
\end{eqnarray}
If the upper limit of the integral defining $\mathcal{I}(a,b)$
is much greater than one (as expected for sub-horizon modes),
such that $\mathcal{I}(a,b) = \mathcal{I}(b,a)$, we then obtain
\begin{eqnarray}\label{}
   g_3 = \int\,\frac{d^3v}{(2\pi)^3}\,\mathcal{I}(v,b)\,
   \mathcal{I}(v,b_2)\,\mathcal{I}(b,b_2)\,
   (\hat\bk\cdot(\bv-{\bf b}))\,\,
   ({\bf u}\cdot(\bv+{\bf b}_2))\,\,
   ((\hat\bk+{\bf u})\cdot({\bf b}-{\bf b}_2))\,.
\label{eq:g3Compact}
\end{eqnarray}
Finally, simple algebraic re-arrangements in
Eq.~(\ref{eq:g3Compact}) then results in the far simpler
expression, in Eq.~(\ref{eqn:biresult}) for $g_3$,
\begin{eqnarray}
g_3 = \int
    \frac{d^3v}{(2\pi)^3}\,H(\bu+\bv,\bv)\,
H(\bv,\hat\bk-\bv)\,H(\hat\bk-\bv,\bu+\bv),
\end{eqnarray}
with $H(a,b) \equiv (b^2-a^2)\mathcal{I}(a,b) \leq 0$. 
This expression is indeed equivalent to Eq.~(59) in 
Ref.~\cite{Jaffe:1993tt}, although it is written in a 
much more compact and simpler way. Note that it is smaller 
by a factor $(2\pi)^3$ than that of Ref.~\cite{Jaffe:1993tt}, 
as a consequence of different conventions.
\end{widetext}

\subsection{Some Integrals and approximations}

Once modes are well inside the horizon, the upper limit for the
integral in Eq.~(\ref{eqn:mathcalI}) is large, and the integral
can then be approximated by
\begin{equation}
    \mathcal{I}(a,b) \equiv \int_0^\infty\, ds\,
    \frac{f(as)f'(bs)}{a^{3/2}b^{1/2}}\,,
\label{eqn:mathcalI2}
\end{equation}
where $f(x)=x^{1/2-\alpha}J_{1+\alpha}$, and $\alpha=2$ for
matter domination.  The integral can be performed analytically;
the result is
\begin{equation}
    \mathcal{I}(a,b) = \begin{cases} \frac{1}{b^3} F(a^2/b^2), &
    \textrm{for $a<b$}, \\ \, & \\
    -\frac{1}{a^3} F(b^2/a^2), & \textrm{for
    $a>b$}, \end{cases}
\end{equation}
where
\begin{eqnarray}
    F(x) &\equiv& \frac{3\sqrt{\pi}}{4^n} \left[ \frac{ x \,
    {}_2F_1 \left( \frac{5}{2},\frac{5}{2}-n;n+2;x \right)} {
    \Gamma(n+2) \Gamma \left( n-\frac{3}{2} \right)}
    \right. \nn \\
    & & \left. -  \frac{ 
    {}_2F_1 \left( \frac{5}{2},\frac{3}{2}-n;n+1;x \right)} {
    \Gamma(n+1) \Gamma \left( n-\frac{1}{2} \right)}
    \right]\,,
\end{eqnarray}
${}_2F_1(w,x;y;z)$ is the hypergeometric function, and
$n=1+\alpha$ is the order of the Bessel function.  While
straightforward to evaluate numerically, this exact solution may
be computationally expensive to evaluate repeatedly.  We
therefore use for our numerical work the approximation,
\begin{equation}
    \mathcal{I}(a,b) \simeq \begin{cases} \frac{-1}{96 b^3}
    \frac{(b/a)^\kappa-1}{(b/a)^\kappa+1}, & \mathrm{if}\ \  a<b,
    \\ \, & \\ \frac{1}{96 a^3} \frac{(a/b)^\kappa-1}{(a/b)^\kappa+1},
    & \mathrm{if}\ \ b<a, \end{cases}
\end{equation}
which provides good agreement with the exact results with
$\kappa=2.5$.


\begin{thebibliography}{99}

\bibitem{inflation}
 A.~H.~Guth and S.~Y.~Pi,
 Phys.\ Rev.\ Lett.\  {\bf 49}, 1110 (1982);
 A.~A.~Starobinsky,
 Phys.\ Lett.\  B {\bf 117}, 175 (1982);
 J.~M.~Bardeen, P.~J.~Steinhardt and M.~S.~Turner,
 Phys.\ Rev.\  D {\bf 28}, 679 (1983).

\bibitem{larger}
 T.~J.~Allen, B.~Grinstein and M.~B.~Wise,
 Phys.\ Lett.\  B {\bf 197}, 66 (1987);
 L.~A.~Kofman and D.~Y.~Pogosian,
 Phys.\ Lett.\  B {\bf 214}, 508 (1988);
 D.~S.~Salopek, J.~R.~Bond and J.~M.~Bardeen,
 Phys.\ Rev.\  D {\bf 40}, 1753 (1989);
 A.~D.~Linde and V.~F.~Mukhanov,
 Phys.\ Rev.\  D {\bf 56}, 535 (1997)
 [arXiv:astro-ph/9610219];
 P.~J.~E.~Peebles, Astrophys.\ J.\  {\bf 510}, 523 (1999)
 [arXiv:astro-ph/9805194];
 P.~J.~E.~Peebles, 
 Astrophys.\ J.\  {\bf 510}, 531 (1999) 
 [arXiv:astro-ph/9805212].

\bibitem{curvaton}
 S.~Mollerach,
 Phys.\ Rev.\  D {\bf 42}, 313 (1990);
 A.~D.~Linde and V.~F.~Mukhanov,
 Phys.\ Rev.\  D {\bf 56}, 535 (1997)
 [arXiv:astro-ph/9610219];
D.~H.~Lyth and D.~Wands,
Phys.\ Lett.\  B {\bf 524}, 5 (2002)
[arXiv:hep-ph/0110002];
 T.~Moroi and T.~Takahashi,
 Phys.\ Lett.\  B {\bf 522}, 215 (2001)
 [Erratum-ibid.\  B {\bf 539}, 303 (2002)]
 [arXiv:hep-ph/0110096];
 D.~H.~Lyth, C.~Ungarelli and D.~Wands,
 Phys.\ Rev.\  D {\bf 67}, 023503 (2003)
 [arXiv:astro-ph/0208055];
 K.~Ichikawa {\it et al.},
 arXiv:0802.4138 [astro-ph];
 K.~Enqvist, S.~Nurmi, O.~Taanila and T.~Takahashi,
 arXiv:0912.4657 [astro-ph.CO];
 K.~Enqvist and T.~Takahashi,
 JCAP {\bf 0912}, 001 (2009)
 [arXiv:0909.5362 [astro-ph.CO]];
 K.~Enqvist and T.~Takahashi,
 JCAP {\bf 0809}, 012 (2008)
 [arXiv:0807.3069 [astro-ph]];
 K.~Enqvist and S.~Nurmi,
 JCAP {\bf 0510}, 013 (2005)
 [arXiv:astro-ph/0508573];
 A.~L.~Erickcek, M.~Kamionkowski and S.~M.~Carroll,
 arXiv:0806.0377 [astro-ph];
 A.~L.~Erickcek, C.~M.~Hirata and M.~Kamionkowski,
 Phys.\ Rev.\  D {\bf 80}, 083507 (2009)
 [arXiv:0907.0705 [astro-ph.CO]].

\bibitem{Dvali:1998pa}
 G.~R.~Dvali and S.~H.~H.~Tye,
 Phys.\ Lett.\  B {\bf 450}, 72 (1999)
 [arXiv:hep-ph/9812483];
 P.~Creminelli,
 JCAP {\bf 0310}, 003 (2003)
 [arXiv:astro-ph/0306122];
 M.~Alishahiha, E.~Silverstein and D.~Tong,
 Phys.\ Rev.\  D {\bf 70}, 123505 (2004)
 [arXiv:hep-th/0404084].

\bibitem{topdefects}
A.~Vilenkin and E.~P.~.S.~Shellard
Cambridge University Press, 1994;
R.~Durrer,
New Astron.\ Rev.\  {\bf 43}, 111 (1999);
R.~Durrer, M.~Kunz and A.~Melchiorri,
Phys.\ Rept.\  {\bf 364}, 1 (2002)
[arXiv:astro-ph/0110348].

\bibitem{endinflation}
 S.~Sarangi and S.~H.~H.~Tye,
 Phys.\ Lett.\  B {\bf 536}, 185 (2002)
 [arXiv:hep-th/0204074];
 N.~T.~Jones, H.~Stoica and S.~H.~H.~Tye,
 Phys.\ Lett.\  B {\bf 563}, 6 (2003)
 [arXiv:hep-th/0303269];
 E.~J.~Copeland, R.~C.~Myers and J.~Polchinski,
 JHEP {\bf 0406}, 013 (2004)
 [arXiv:hep-th/0312067];
 G.~Dvali and A.~Vilenkin,
 JCAP {\bf 0403}, 010 (2004)
 [arXiv:hep-th/0312007];
R.~Jeannerot, J.~Rocher and M.~Sakellariadou,
Phys.\ Rev.\  D {\bf 68}, 103514 (2003)
[arXiv:hep-ph/0308134].
 L.~Kofman, A.~D.~Linde and A.~A.~Starobinsky,
 Phys.\ Rev.\ Lett.\  {\bf 76}, 1011 (1996)
 [arXiv:hep-th/9510119];
G.~N.~Felder, J.~Garcia-Bellido, P.~B.~Greene, L.~Kofman, A.~D.~Linde and 
I.~Tkachev,
Phys.\ Rev.\ Lett.\  {\bf 87}, 011601 (2001)
[arXiv:hep-ph/0012142];
G.~N.~Felder, L.~Kofman and A.~D.~Linde,
Phys.\ Rev.\  D {\bf 64}, 123517 (2001)
[arXiv:hep-th/0106179].

\bibitem{Turok:1991qq}
 N.~Turok and D.~N.~Spergel,
 Phys.\ Rev.\ Lett.\  {\bf 66}, 3093 (1991);
 M.~Kunz and R.~Durrer,
 Phys.\ Rev.\  D {\bf 55}, 4516 (1997)
 [arXiv:astro-ph/9612202];
\bibitem{Durrer:1998rw}
 R.~Durrer, M.~Kunz and A.~Melchiorri,
 Phys.\ Rev.\  D {\bf 59}, 123005 (1999)
 [arXiv:astro-ph/9811174].

\bibitem{Bevis:2004wk}
 N.~Bevis, M.~Hindmarsh and M.~Kunz,
 Phys.\ Rev.\  D {\bf 70}, 043508 (2004)
 [arXiv:astro-ph/0403029].

\bibitem{Crotty:2003rz}
P.~Crotty, J.~Garcia-Bellido, J.~Lesgourgues and A.~Riazuelo,
Phys.\ Rev.\ Lett.\  {\bf 91}, 171301 (2003)
[arXiv:astro-ph/0306286];
M.~Beltran, J.~Garcia-Bellido, J.~Lesgourgues and M.~Viel,
Phys.\ Rev.\  D {\bf 72}, 103515 (2005)
[arXiv:astro-ph/0509209].

\bibitem{CMBtexture}
 M.~Cruz, N.~Turok, P.~Vielva, E.~Martinez-Gonzalez and M.~Hobson,
 Science {\bf 318}, 1612 (2007)
 [arXiv:0710.5737 [astro-ph]];
 C.~L.~Bennett {\it et al.},
 arXiv:1001.4758 [astro-ph.CO].


\bibitem{NGReview}
 N.~Bartolo {\it et al.},
 Phys.\ Rept.\  {\bf 402}, 103 (2004)
 [arXiv:astro-ph/0406398].

\bibitem{Bernardeau:2001qr}
 F.~Bernardeau, S.~Colombi, E.~Gaztanaga and R.~Scoccimarro,
 Phys.\ Rept.\  {\bf 367}, 1 (2002)
 [arXiv:astro-ph/0112551].

\bibitem{localmodel}
 T.~Falk, R.~Rangarajan and M.~Srednicki,
 Astrophys.\ J.\  {\bf 403}, L1 (1993)
 [arXiv:astro-ph/9208001];
 A.~Gangui {\it et al.},
 Astrophys.\ J.\  {\bf 430}, 447 (1994)
 [arXiv:astro-ph/9312033];
 A.~Gangui,
 Phys.\ Rev.\  D {\bf 50}, 3684 (1994)
 [arXiv:astro-ph/9406014].

\bibitem{Wang:2000}
 L.~M.~Wang and M.~Kamionkowski,
 Phys.\ Rev.\  D {\bf 61}, 063504 (2000)
 [arXiv:astro-ph/9907431].

\bibitem{Maldacena:2002vr}
 J.~M.~Maldacena,
 JHEP {\bf 0305}, 013 (2003)
 [arXiv:astro-ph/0210603].

\bibitem{Equil}
 D.~Babich, P.~Creminelli and M.~Zaldarriaga,
 JCAP {\bf 0408}, 009 (2004)
 [arXiv:astro-ph/0405356];
 P.~Creminelli {\it et al.},
 JCAP {\bf 0605}, 004 (2006)
 [arXiv:astro-ph/0509029];
 P.~Creminelli {\it et al.},
 JCAP {\bf 0703}, 005 (2007)
 [arXiv:astro-ph/0610600].

\bibitem{Hannestad:2009yx}
 S.~Hannestad, T.~Haugboelle, P.~R.~Jarnhus and M.~S.~Sloth,
 arXiv:0912.3527 [hep-ph].

\bibitem{Luo:1993xx}
 X.~c.~Luo,
 Astrophys.\ J.\  {\bf 427}, L71 (1994)
 [arXiv:astro-ph/9312004];
 L.~Verde {\it et al.},
 Mon.\ Not.\ Roy.\ Astron.\ Soc.\  {\bf 313}, L141 (2000)
 [arXiv:astro-ph/9906301];
 E.~Komatsu and D.~N.~Spergel,
 Phys.\ Rev.\  D {\bf 63}, 063002 (2001)
 [arXiv:astro-ph/0005036].

\bibitem{LSS}
 P.~Coles {\it et al.}, 
 Mon.\ Not.\ Roy.\ Astron.\ Soc.\  {\bf 264}, 749 (1993)
 [arXiv:astro-ph/9302015];
 X.~c.~Luo and D.~N.~Schramm,
 Astrophys.\ J.\  {\bf 408}, 33 (1993);
 E.~Lokas {\it et al.}, Mon.\ Not.\ Roy.\ Astron.\ Soc.\  {\bf
 274}, 730 (1995) 
 [arXiv:astro-ph/9407095];
 M.~J.~Chodorowski and F.~R.~Bouchet, Mon.\ Not.\ Roy.\
 Astron.\ Soc.\  {\bf 279}, 557 (1996) 
 [arXiv:astro-ph/9507038];
 A.~J.~Stirling and J.~A.~Peacock, Mon.\ Not.\ Roy.\ Astron.\
 Soc.\  {\bf 283}, 99 (1996) 
 [arXiv:astro-ph/9608101];
 R.~Durrer {\it et al.},
 Phys.\ Rev.\  D {\bf 62}, 021301 (2000)
 [arXiv:astro-ph/0005087].
 L.~Verde and A.~F.~Heavens,
 Astrophys.\ J.\  {\bf 553}, 14 (2001)
 [arXiv:astro-ph/0101143];
 A.~Buchalter and M.~Kamionkowski,
 Astrophys.\ J.\  {\bf 521}, 1 (1999)
 [arXiv:astro-ph/9903462];
 A.~Buchalter, M.~Kamionkowski and A.~H.~Jaffe,
 Astrophys.\ J.\  {\bf 530}, 36 (2000)
 [arXiv:astro-ph/9903486].

\bibitem{abundances}
 W.~A.~Chiu, J.~P.~Ostriker and M.~A.~Strauss,
 Astrophys.\ J.\  {\bf 494}, 479 (1998)
 [arXiv:astro-ph/9708250];
 J.~Robinson, E.~Gawiser and J.~Silk,
 Astrophys.\ J.\  {\bf 532}, 1 (2000)
 [arXiv:astro-ph/9906156];
 J.~A.~Willick, Astrophys.\ J.\  {\bf 530}, 80 (2000)
 [arXiv:astro-ph/9904367];
 L.~Verde {\it et al.},
 Mon.\ Not.\ Roy.\ Astron.\ Soc.\  {\bf 321}, L7 (2001)
 [arXiv:astro-ph/0007426];
 N.~N.~Weinberg and M.~Kamionkowski,
 Mon.\ Not.\ Roy.\ Astron.\ Soc.\  {\bf 341}, 251 (2003)
 [arXiv:astro-ph/0210134];
 S.~Matarrese, L.~Verde and R.~Jimenez,
 Astrophys.\ J.\  {\bf 541}, 10 (2000)
 [arXiv:astro-ph/0001366];
 L.~Verde {\it et al.},
 Mon.\ Not.\ Roy.\ Astron.\ Soc.\  {\bf 325}, 412 (2001)
 [arXiv:astro-ph/0011180];
 M.~LoVerde {\it et al.},
 JCAP {\bf 0804}, 014 (2008)
 [arXiv:0711.4126 [astro-ph]].

\bibitem{Kamionkowski:2008sr}
 M.~Kamionkowski, L.~Verde and R.~Jimenez,
 JCAP {\bf 0901}, 010 (2009)
 [arXiv:0809.0506 [astro-ph]].

\bibitem{Dalal:2007cu}
 N.~Dalal {\it et al.},
 Phys.\ Rev.\  D {\bf 77}, 123514 (2008)
 [arXiv:0710.4560 [astro-ph]];
 S.~Matarrese and L.~Verde,
 Astrophys.\ J.\  {\bf 677}, L77 (2008)
 [arXiv:0801.4826 [astro-ph]];
 A.~Slosar {\it et al.},
 arXiv:0805.3580 [astro-ph];
 C.~Carbone, L.~Verde and S.~Matarrese,
 arXiv:0806.1950 [astro-ph].

\bibitem{Jaffe:1993tt}
 A.~H.~Jaffe,
 Phys.\ Rev.\  D {\bf 49}, 3893 (1994)
 [arXiv:astro-ph/9311023].

\bibitem{DefectBITRISPECTRA}
 A.~Silvestri and M.~Trodden,
 Phys.\ Rev.\ Lett.\  {\bf 103}, 251301 (2009)
 [arXiv:0811.2176 [astro-ph]];
 M.~Hindmarsh, C.~Ringeval and T.~Suyama,
 Phys.\ Rev.\  D {\bf 80}, 083501 (2009)
 [arXiv:0908.0432 [astro-ph.CO]];
 M.~Hindmarsh, C.~Ringeval and T.~Suyama,
 arXiv:0911.1241 [astro-ph.CO];
 D.~M.~Regan and E.~P.~S.~Shellard,
 arXiv:0911.2491 [astro-ph.CO].

\bibitem{Kamionkowski:1992mf}
 M.~Kamionkowski and J.~March-Russell,
 Phys.\ Lett.\  B {\bf 282}, 137 (1992)
 [arXiv:hep-th/9202003];
 M.~Kamionkowski and J.~March-Russell,
 Phys.\ Rev.\ Lett.\  {\bf 69}, 1485 (1992)
 [arXiv:hep-th/9201063];
 R.~Holman, S.~D.~H.~Hsu, T.~W.~Kephart, E.~W.~Kolb, R.~Watkins and L.~M.~Widrow,
 Phys.\ Lett.\  B {\bf 282}, 132 (1992)
 [arXiv:hep-ph/9203206];
 R.~Holman, S.~D.~H.~Hsu, E.~W.~Kolb, R.~Watkins and L.~M.~Widrow,
 Phys.\ Rev.\ Lett.\  {\bf 69}, 1489 (1992).

\bibitem{Barr:1992qq}
 S.~M.~Barr and D.~Seckel,
 Phys.\ Rev.\  D {\bf 46}, 539 (1992).

\bibitem{Veeraraghavan:1990yd}
 S.~Veeraraghavan and A.~Stebbins,
 Astrophys.\ J. {\bf 365}, 37 (1990).

\bibitem{Pen:1997ae}
 U.~L.~Pen, U.~Seljak and N.~Turok,
 Phys.\ Rev.\ Lett.\  {\bf 79}, 1611 (1997)
 [arXiv:astro-ph/9704165].

\bibitem{Turok:1997gj}
 N.~Turok, U.~L.~Pen and U.~Seljak,
 Phys.\ Rev.\  D {\bf 58}, 023506 (1998)
 [arXiv:astro-ph/9706250].

\bibitem{Pen:1993nx}
 U.~L.~Pen, D.~N.~Spergel and N.~Turok,
 Phys.\ Rev.\  D {\bf 49}, 692 (1994).

\bibitem{Creminelli:2005hu}
 P.~Creminelli, A.~Nicolis, L.~Senatore, M.~Tegmark and M.~Zaldarriaga,
 JCAP {\bf 0605}, 004 (2006)
 [arXiv:astro-ph/0509029].

\bibitem{limits}
 E.~Komatsu {\it et al.}  [WMAP Collaboration],
 Astrophys.\ J.\ Suppl.\  {\bf 148}, 119 (2003)
 [arXiv:astro-ph/0302223];
 A.~P.~S.~Yadav and B.~D.~Wandelt,
 Phys.\ Rev.\ Lett.\  {\bf 100}, 181301 (2008)
 [arXiv:0712.1148 [astro-ph]];
 E.~Komatsu {\it et al.}  [WMAP Collaboration],
 Astrophys.\ J.\ Suppl.\  {\bf 180}, 330 (2009)
 [arXiv:0803.0547 [astro-ph]].

\bibitem{Komatsu:2010fb}
 E.~Komatsu {\it et al.},
 arXiv:1001.4538 [astro-ph.CO].

\bibitem{Cooray:2008xz}
 A.~Cooray, D.~Sarkar and P.~Serra,
 Phys.\ Rev.\  D {\bf 77}, 123006 (2008)
 [arXiv:0803.4194 [astro-ph]].

\bibitem{Seljak:1997ii}
 U.~Seljak, U.~L.~Pen and N.~Turok,
 Phys.\ Rev.\ Lett.\  {\bf 79}, 1615 (1997)
 [arXiv:astro-ph/9704231].

\bibitem{Kamionkowski:1996ks}
 M.~Kamionkowski, A.~Kosowsky and A.~Stebbins,
 Phys.\ Rev.\  D {\bf 55}, 7368 (1997)
 [arXiv:astro-ph/9611125];
 M.~Kamionkowski, A.~Kosowsky and A.~Stebbins,
 Phys.\ Rev.\ Lett.\  {\bf 78}, 2058 (1997)
 [arXiv:astro-ph/9609132];
 M.~Zaldarriaga and U.~Seljak,
 Phys.\ Rev.\  D {\bf 55}, 1830 (1997)
 [arXiv:astro-ph/9609170].
 U.~Seljak and M.~Zaldarriaga,
 Phys.\ Rev.\ Lett.\  {\bf 78}, 2054 (1997)
 [arXiv:astro-ph/9609169].

\bibitem{Baumann:2009mq}
D.~Baumann and M.~Zaldarriaga,
JCAP {\bf 0906}, 013 (2009)
[arXiv:0901.0958 [astro-ph.CO]];
P.~Adshead and E.~A.~Lim,
arXiv:0912.1615 [astro-ph.CO].

\bibitem{Krauss:1991qu}
 L.~M.~Krauss,
 Phys.\ Lett.\  B {\bf 284}, 229 (1992);
 E.~Fenu, D.~G.~Figueroa, R.~Durrer and J.~Garcia-Bellido,
 JCAP {\bf 0910}, 005 (2009)
 [arXiv:0908.0425 [astro-ph.CO]];
 K.~Jones-Smith, L.~M.~Krauss and H.~Mathur,
 Phys.\ Rev.\ Lett.\  {\bf 100}, 131302 (2008)
 [arXiv:0712.0778 [astro-ph]].

\end{thebibliography}
\end{document}